\documentstyle[multicol,prl,aps,epsf]{revtex}
\begin{document}

\title{The influence of critical behavior on the spin glass 
phase}

\author{Hemant Bokil}
\address{Abdus Salam ICTP, Strada Costiera 11, 34100 Trieste, 
Italy}
\author{Barbara Drossel}
\address{School of Physics and Astronomy, Raymond and 
Beverley Sackler Faculty of Exact Sciences, Tel Aviv 
University, Tel Aviv 69978, Israel}
\author{M. A. Moore}
\address{Department of Physics, University of Manchester, 
Manchester M13 9PL, U.K.}
\date{\today} 
\maketitle

\begin{abstract}
  We have argued in recent papers that the Monte Carlo 
results for the
  equilibrium properties of the Edwards-Anderson spin glass 
in three
  dimensions, which had been interpreted earlier as providing 
evidence
  for replica symmetry breaking, can be explained quite simply within 
the
  droplet model once finite size effects and proximity to the 
critical
  point are taken into account.  In this paper we show that 
similar
  considerations are sufficient to explain the Monte Carlo 
data in
  four dimensions. In particular, we study the Parisi overlap 
and the
  link overlap for the four-dimensional Ising spin glass in 
the
  Migdal-Kadanoff approximation. Similar to what is seen in 
three
  dimensions, we find that temperatures well below those 
studied in
  the Monte Carlo simulations have to be reached before the 
droplet model
  predictions become apparent. We also show that the double-peak
  structure of the link overlap distribution function is 
related to
  the difference between domain-wall excitations that cross 
the entire
  system and droplet excitations that are confined to a 
smaller
  region.

  \pacs{PACS numbers:
75.50.Lk Spin glasses}
\end{abstract}
\begin{multicols}{2}
  
\section{Introduction}


  Despite over two decades of work, the controversy 
concerning the
  nature of the ordered phase of short range Ising spin 
glasses
  continues.  For a few years, Monte Carlo simulations 
appeared to be
  providing evidence for replica symmetry breaking (RSB) in 
these
  systems\cite{rby,marinari}. However, recent developments 
have
  cast doubt on this interpretation of the Monte Carlo data.
  In a series of papers on the Ising spin glass within the
  Migdal-Kadanoff approximation (MKA), we showed that the 
equilibrium
  Monte Carlo data in three dimensions that had been 
interpreted in
  the past as giving evidence for RSB can actually be 
interpreted
  quite easily within the droplet picture, with apparent RSB 
effects
  being attributed to a crossover between critical behavior 
and the
  asymptotic droplet-like behavior for small system
  sizes\cite{us1,us2,us3,us4}.  We also showed that system 
sizes
  well beyond the reach of current simulations would probably 
be 
  required in order to unambiguously see droplet-like 
behavior.
  The finding that the critical-point
  effects can still be felt at temperatures lower than those
  accessible by Monte Carlo simulations is supported by the 
Monte
  Carlo simulations of Berg and Janke\cite{berg} who found 
critical
  scaling working reasonably well down to $T=0.8T_c$ for 
system sizes
  upto $L=8$ in three dimensions. The zero temperature study 
  of Pallasini and Young\cite{py} also suggests that the 
ground-
  state structure of three-dimensional Edwards-Anderson model 
  is well described by droplet theory, though the existence 
of low 
  energy excitations not included in
  the conventional droplet theory remains an open question. 
Thus,
  while puzzles do remain, the weight of the evidence seems 
to be
  shifting towards a droplet-like description of the ordered 
phase in
  short range Ising spin glasses.
  
  However, it is expected that critical point effects are 
less
  dominant in four dimensions than in three dimensions.  Our 
aim in
  this paper is to quantify the extent of critical point 
effects in
  the low temperature phase of the four-dimensional 
Edwards-Anderson
  spin glass. We do this by providing results for the 
four-dimensional
  Ising spin glass in the MKA and compare these with existing 
Monte
  Carlo work.  In particular, we study the Parisi overlap 
function and
  the link overlap function for system sizes up to $L=16$ and
  temperatures as low as $T=0.16T_c$.  We find that for 
system sizes
  and temperatures comparable to those of the Monte Carlo 
simulations,
  the Parisi overlap distribution shows also in MKA the
  sample-to-sample fluctations and the stationary behavior at 
small
  overlap values, that are normally attributed to RSB. It is 
only for
  larger system sizes (or for lower temperatures), that the 
asymptotic
  droplet-like behavior becomes apparent. For the link 
overlap, we
  find similar double-peaked curves as those found in 
Monte-Carlo
  simulations. This double peak structure is expected on 
quite general
  grounds independent of the nature of the low temperature 
phase.
  However, we show that two peaks in the link overlap in MKA 
occur
  because of a difference between domain-wall excitations 
(which cross
  the entire system) and droplet excitations (which do not 
cross the
  entire system). We argue that for small system sizes, the 
effect of
  domain walls increases with increasing dimension, making it
  necessary to go very far below $T_c$ to see the asymptotic 
droplet
  behavior.
  
  This paper is organized as follows: in section 
\ref{definitions}, we
  define the quantities discussed in this paper, and the 
droplet-model
  predictions for their behavior.  In section \ref{MKA}, we 
describe
  the MKA, and our numerical methods of evaluating the 
overlap
  distribution.  In section \ref{Parisi}, we present our 
numerical
  results for the Parisi overlap distribution, and compare to
  Monte-Carlo data. The following section studies the link 
overlap
  distribution.  Finally, section \ref{conclusions} contains 
the
  concluding remarks, including some on the effects of 
critical behavior on the
  dynamics in the spin glass phase. Again we suspect that 
arguments which have
been advanced against the droplet picture on the basis of 
dynamical studies
have failed to take into account the effects arising from 
proximity to the
critical point.

\section{Definitions and Scaling Laws}
\label{definitions}

The Edwards-Anderson spin glass in the absence of an external 
magnetic field is
defined by
the Hamiltonian
$$H=-\sum_{\langle i,j\rangle} J_{ij} S_iS_j,$$ where the 
Ising spins can
take the values $\pm 1$, and the nearest-neighbor couplings 
$J_{ij}$
are independent from each other and Gaussian distributed with 
a
standard deviation $J$. 

It has proven useful to consider two identical copies 
(replicas) of
the system, and to measure overlaps between them. This gives
information about the structure of the low-temperature phase, 
in
particular about the number of pure states. The quantities 
considered in this 
paper are the Parisi overlap function $P(q,L)$ and the
link overlap function $P(q_l,L)$. They are defined by
\begin{equation} 
P(q, L) = \left[\left<\delta \left(\sum_{\langle i j\rangle} 
\frac{{S_i^{(1)}S_i^{(2)} + S_j^{(1)}S_j^{(2)}}} 
{2N_L} - q
\right)\right>\right],
\label{p}
\end{equation}
and 
\begin{equation} 
P(q_l, L)=  \left[\left<\delta\left(\sum_{\langle i j\rangle} 
\frac{S_i^{(1)}S_i^{(2)}S_j^{(1)}S_j^{(2)}}
{N_L}-q_l
\right)\right>\right].
\label{pl}
\end{equation}
Here, the superscripts $(1)$ and $(2)$ denote the two 
replicas of the
system, $N_L$ is the number of bonds, and $\langle 
...\rangle$ and
$\left[...\right]$ denote the thermodynamic and disorder 
average
respectively.  We use $P(q, L)$ and $P(q_l,L)$ to denote the 
overlap functions
for a finite system of size $L$, reserving the more standard 
notation
$P(q)$ and $P(q_l)$ for the limit $\lim_{L\to \infty} P(q, 
L)$
and $\lim_{L\to \infty} P(q_l, L)$.

In the mean-field RSB picture, $P(q)$ is nonzero in the spin 
glass
phase in the entire interval $[-q_{EA},q_{EA}]$, while it is 
composed
only of two delta functions at $\pm q_{EA}$ in the droplet 
picture.
Similarly, $P(q_l)$ is nonzero over a finite interval 
$[q_l^{min},
q_l^{max}]$ in mean-field theory, while it is a 
delta-function within
the droplet picture.

Much of the evidence for RSB for three- and four-dimensional 
systems
comes from observing a stationary $P(q=0,L)$ for system sizes 
that are
generally smaller than 20 in 3D and smaller than 10 in 4D, 
and at
temperatures of the order of $0.7 T_c$.  However, even within 
the
droplet picture one expects to see a stationary $P(q=0, L)$ 
for a
certain range of system sizes and temperatures.  The reason 
is that at
$T_c$ the overlap distribution $P(q,L)$ obeys the scaling law
\begin{equation}
P(q,L)=L^{\beta/\nu} \tilde P(q L^{\beta/\nu}),
\label{scaling}
\end{equation}
$\beta$ being the order parameter critical exponent, and 
$\nu$ the
correlation length exponent. Above the lower critical 
dimension (which
is smaller than 3), $\beta/\nu$ is positive, leading to an 
increase
$P(q=0,L)$ as a function of $L$ (at $T=T_c$). On the other 
hand, for
$T\ll T_c$, the droplet model predicts a decay $$P(q=0,L) 
\sim
1/L^\theta$$
on length scales larger than the (temperature--dependent)
correlation length $\xi$, $\theta$ being the scaling exponent 
of the
coupling strength $J$. A few words are in order here on what 
we mean
by the correlation length.
In the spin glass phase, all correlation functions fall off
as a power law at large distances. However, within the 
droplet
model, this is true only asymptotically, and the general
form of the correlation function for
two spins a distance $r$ apart, at a temperature $T\le T_c$, 
is 
$\sim r^{-\theta} f(r/\xi)$
where $k_B$ is the Boltzmann constant and $f$ is a scaling 
function.
Thus, for $r\le \xi$ there are
corrections to the algebraic long-distance behavior and 
the above expression defines
the temperature-dependent correlation length.
Note that for $T\to T_c$ this correlation length is 
expected to diverge with the exponent $\nu$.

Thus, for temperatures not too far below $T_c$, one can 
expect an
almost stationary $P(q=0, L)$ for a certain range of system 
sizes. In
three dimensions both $\beta/\nu\simeq0.3$\cite{berg} and
$\theta\simeq0.17$\cite{bm} are rather small, this apparent
stationarity may persist over a considerable range of system 
sizes
$L$. However, in four dimensions, $\beta/\nu\simeq 0.85$ 
\cite{mar99}
and $\theta\simeq 0.65$ \cite{hartmann} and one would expect 
the
crossover region to be smaller. In the present paper we shall
investigate these crossover effects in four dimensions by 
studying
$P(q,L)$ for the Edwards-Anderson spin glass within the MKA. 
It turns
out that they are surprisingly persistent even at  low 
temperatures,
due to the presence of domain walls.

Monte-Carlo simulations of the link overlap distribution show 
a
nontrivial shape with shoulders or even a double peak, which 
seems to
be incompatible with the droplet picture, where the 
distribution
should tend towards a delta-function. For sufficiently low
temperatures and large length scales, the droplet picture 
predicts
that the width of the link overlap distribution scales as 
\cite{us4}
$$\Delta q_l \sim \sqrt{kT} L^{d_s-d-\theta/2}\,,$$
where $d_s$ is the
fractal dimension of a domain wall.  Below, we will show that 
the
nontrivial shape and the double peak reported from 
Monte-Carlo
simulations are also found in MKA in four dimensions, and we 
will
present strong evidence that it is due to the different 
nature of
droplet and domain wall excitations. As the weight of domain 
walls
becomes negligible in the thermodynamic limit, the droplet 
picture is
regained on large scales.

\section{Migdal-Kadanoff approximation}
\label{MKA}

\begin{figure}
\centerline{
\epsfysize=0.15\columnwidth{\epsfbox{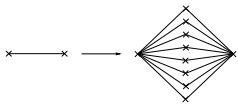}}}
\narrowtext{\caption{Construction of a hierarchical 
lattice.}\label{fig1}}
\end{figure}

The Migdal-Kadanoff approximation (MKA) is a real-space 
renormalization group the gives approximate recursion 
relations for the various coupling constants.  Evaluating a 
thermodynamic quantity in MKA
in $d$ dimensions is equivalent to evaluating it on an
hierarchical lattice that is constructed iteratively by 
replacing each
bond by $2^d$ bonds, as indicated in Fig.~\ref{fig1}. The 
total number of bonds
after $I$ iterations is $2^{dI}$. $I=1$, the smallest 
non-trivial system
that can be studied, corresponds to a system linear
dimension $L=2$, $I=2$ corresponds to $L=4$, $I=3$ 
corresponds to $L=8$
and so on. Note that the number of bonds on hierarchical 
lattice after
$I$ iterations is the same as the number of 
 sites of a $d$-dimensional lattice of size $L=2^I$.
Thermodynamic quantities are then evaluated iteratively by 
tracing
over the spins on the highest level of the hierarchy, until 
the
lowest level is reached and the trace over the remaining two 
spins is
calculated \cite{southern77}. This procedure generates new
effective couplings, which have to be included in the 
recursion
relations.

In
\cite{gardner84}, it was proved that in the limit of 
infinitely many
dimensions (and in an expansion away from infinite 
dimensions) the MKA
reproduces the results of the droplet picture.

As was discussed in \cite{us1}, the calculation of $P(q,L)$ 
is
made easier by first calculating its 
Fourier transform $F(y,L)$, which is given by
\begin{equation}
F(y,L)=\left[\left< \exp[iy\sum_{\langle i j\rangle}
{(S_i^{(1)}S_i^{(2)}+S_j^{(1)}S_j^{(2)})\over {2N_L}}] 
\right>\right] .
\end{equation}
The recursion relations for $F(y,L)$ involve two-
and four-spin terms, and can easily be evaluated numerically 
because all
terms are now in an exponential. Having calculated $F(y)$ one 
can
then invert the Fourier transform to get $P(q,L)$. 

Similarly, $P(q_l,L)$ is calculated by first evaluating 
\begin{equation}
F(y_l,L)=\left[\left< \exp[iy_l\sum_{\langle i j\rangle}
{(S_i^{(1)}S_i^{(2)}S_j^{(1)}S_j^{(2)})\over {N_L}}] 
\right>\right] .
\end{equation}

\begin{figure}
  \centerline{
    \epsfysize=0.7\columnwidth{\epsfbox{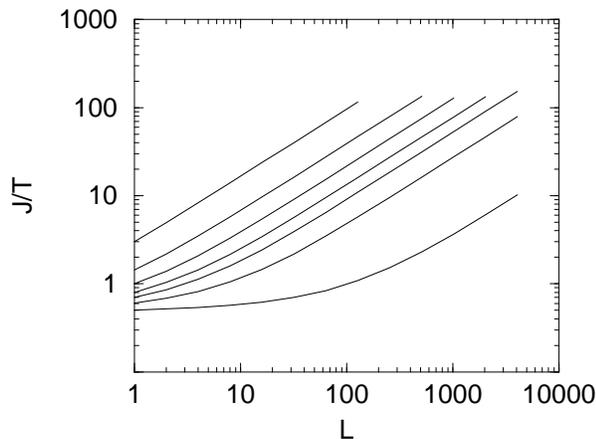}}}
  \narrowtext{\caption{Flow of the coupling strength $J$ in 
MKA. The
      curves correspond to $T/T_c=$0.96, 0.8, 0.68, 0.6, 
0.48, 0.33,
      0.16 (from bottom to top).  The correlation lengths, 
where the
      slope has reached $90\%$ of the asymptotic slope, are 
960, 47,
      24, 15, 8, 3, 1.}\label{flow} }
\end{figure}

Before presenting our numerical results for the Parisi 
overlap and the
link overlap, let us discuss the flow of the coupling 
constant $J$ in
the low-temperature phase, as obtained in MKA. In order to 
obtain this
flow, we iterated the MKA recursion relation on a set of 
$10^6$ bonds.
At each iteration, each of the new set of $10^6$ bonds was 
generated
by randomly choosing 16 bonds from the old set and taking the 
trace
over the inner spins (with a bond arrangement as in 
Fig.~\ref{fig1}).
Figure \ref{flow} shows $J/T$ as function of $L$ for 
different initial
values of the coupling strength.  The critical point is at 
$T_c \simeq
2.1 J$. The first curve begins at $J/T=0.5$, which is close 
to the
critical point, and it reaches the low-temperature behavior 
only at
lengths around 1000.  For an initial $J/T=0.7$, the 
asymptotic slope
is already reached at $L$ around 40, and for $J/T=3.0$, which
corresponds to $T\simeq 0.16T_c$ the entire curve shows the 
asymptotic
slope.  The asymptotic slope is identical to the 
above-mentioned
exponent $\theta$ and has the value $\theta \simeq 0.75$. In 
contrast to
$d=3$ \cite{us4}, we did not succeed in fitting the crossover 
regime
by doing an expansion around the zero-temperature fixed 
point. The
reason is that dimension 4 is too far above the lower 
critical
dimension, such that the critical temperature is not small. 

Note that for each temperature the length scale beyond which 
the flows
of the coupling constants show the asymptotic behavior yields 
one
estimate for the correlation length mentioned above. We have
considered the flow to be in the asymptotic regime when its 
slope was
within 90\% of its asymptotic value. However, this estimate 
is
specific to the flows of the coupling constant, and other
quantities may show their asymptotic behavior later.
In fact, as we shall see below, the convergence of the
overlap distributions is much slower than that of the 
couplings, and
we will have to give reasons for this.

\section{The Parisi overlap}
\label{Parisi}

\begin{figure}
\centerline{
\epsfysize=0.7\columnwidth{\epsfbox{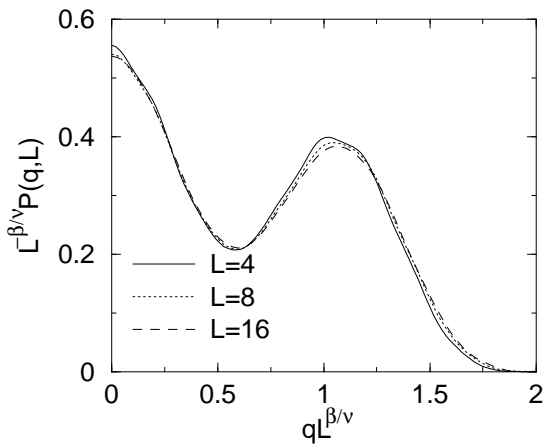}}}
\narrowtext{\caption{Scaling collapse of $P(q,L)$ at $T=T_c$, 
with 
$\beta/\nu\simeq 0.64$. As $P(q,L)=P(-q,L)$, only the part 
$q\ge 0$ is shown.
For each system size, we averaged over at least 5000 
samples.}\label{fig2} 
}
\end{figure}

We now discuss our results for the Parisi overlap.  First, 
let us
briefly describe the critical behavior. Fig.~\ref{fig2} shows 
a
scaling plot for $P(q,L)$ for $L=4,8,16$ at $T=T_c\simeq 
2.1J$.  We
find a good data collapse if we use the value 
$\beta/\nu=0.64$, thus
confirming the finite-size scaling ansatz Eq.~\ref{scaling}.

\begin{figure}
\centerline{
\epsfysize=0.7\columnwidth{\epsfbox{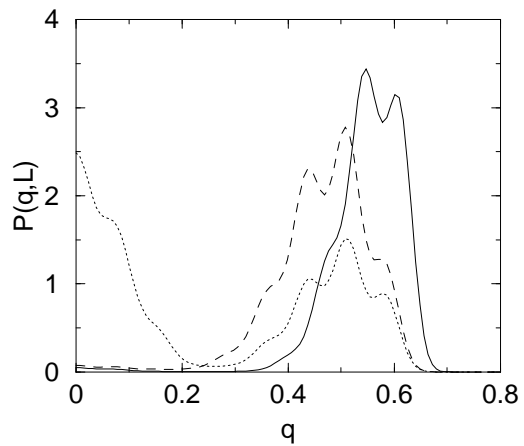}}}
\narrowtext{\caption{$P(q,L)$ for three different samples at 
$T=0.5T_c$
and $L=8$.}\label{samples} 
}
\end{figure}

We next move on to the  low-temperature phase. 
In Fig.~\ref{samples} we show $P(q,L)$ at $T=0.5T_c$
and $L=8$ for three different samples. 
As one can see there are substantial differences between the 
samples.
This sensitivity to samples for system sizes around 10 is in
\cite{mar99} interpreted as evidence for RSB. In our case, 
where we
know that the droplet model is exact, it has to be considered 
a finite
size effect. Note that we have not chosen the three samples 
in any
particular manner. By comparing to the curves obtained for 
$L=16$ (not
shown), we can even see the trend to an increasing number of 
peaks,
just as in \cite{mar99}.  Thus, one feature commonly 
associated with
RSB is certainly present in within the MKA for temperatures 
and system
sizes comparable to those studied in simulations.

Let us now focus on the behavior of $P(q=0,L)$ for
different system sizes and temperatures. But before 
exhibiting
our own data, we discuss 
the Monte Carlo data of Reger, Bhatt and Young\cite{rby}
who were the first to study $P(q=0,L)$ for the 
Edwards-Anderson spin glass.
They studied system sizes $L=2,3,4,5,6$ at temperatures
down to $T=0.68T_c$. At $T=T_c$ they found the expected 
critical
scaling, $P(q=0)\simeq L^{\beta/\nu}$ 
with $\beta/\nu\simeq 0.75$.
Then, as the temperature was lowered, the curves
for $P(q=0)$ as a function of $L$ showed a downward curvature 
for the
largest system sizes,
which they interpreted as the beginning of the crossover 
between
critical behavior and the low temperature behavior. At 
$T=0.8T_c$,
$P(q=0)$ seemed to be roughly constant or decreasing slowly. 
However,
the striking part of their data was that at $T=0.68T_c$ they
found that $P(q=0,L)$ initially decreased as a function of 
system
size for $L=2,3,4$ and then saturated for $L=4,5,6$. They
interpreted this as suggestive of RSB. They admitted however
that other explanations are possible. 

The most recent Monte-Carlo simulation data for the 4d Ising 
spin
glass are those in \cite{mar99}. These authors focus on 
$T\simeq
0.6T_c$, and they find an essentially stationary $P(q=0,L)$ 
for system
sizes up to the largest simulated size $L=10$. They argue, 
that
stationarity over such a large range of $L$ values is most 
naturally
interpreted as evidence for RSB. However, as can be seen from
Fig.~\ref{fig2}, the correlation length is of the order of 16 
for
these temperatures and therefore comparable to the system 
size.

\begin{figure}
\centerline{\epsfysize=0.7\columnwidth{\epsfbox{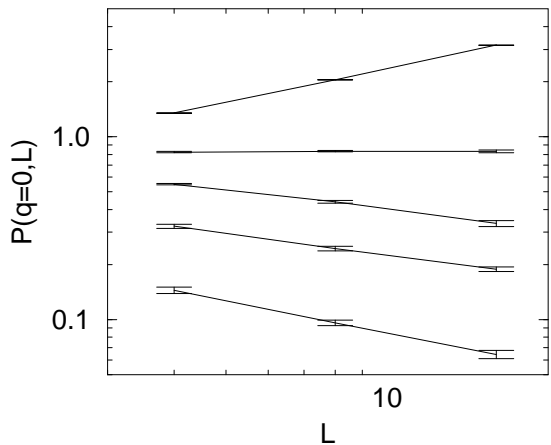}}}
\narrowtext{\caption{$P(q=0,L)$ at $T=T_c$, $0.68T_c$, 
$0.5T_c$, $0.33T_c$, $0.16 T_c$ for
L=4,8,16. The error bars indicate the standard deviation of 
the values. All data were obtained by averaging at least over 
5000 samples.}\label{fig4} 
} 
\end{figure}

In Fig.\ref{fig4}, we show the MKA data for $P(q=0,L)$. We 
have
calculated $P(q=0,L)$ for system sizes $L=4$,$8$,$16$ at 
temperatures
$T=T_c$, $0.68T_c$, $0.33T_c$, and $0.16T_c$.  At $T=T_c$, 
$P(q=0,L)$
grows as $L^{\beta/\nu}$ with $\beta/\nu\simeq 0.64$, in 
agreement
with Fig.\ref{fig2}.  At $T=0.68T_c$ (the lowest temperature 
studied
in \cite{rby}, and not far from the lowest temperature 
studied by
\cite{mar99}), we do not see a clear decrease even for 
$L=16$.  The
curve for $P(q=0)$ looks more or less flat, though one could 
say that
there is slight increase between $L=4$ and $L=8$ and a slight 
decrease
between $L=8$ and $L=16$.  This flat behavior is similar to 
what was
found in \cite{rby} and \cite{mar99}.  The deviation of the 
$L=2$ and
$L=3$ data from the flat curve in \cite{rby} can probably be 
ascribed
to artifacts at very small system sizes, which are also found
elsewhere \cite{bhatt88}.  For lower temperatures, where the
correlation length is smaller than the system size, there is 
a clear
decrease of $P(q=0)$ although the decrease is not asympotic 
even at a
temperatures as low as $T_c/6$.
We conclude that the observed stationarity of $P(q=0,L)$ in
Monte-Carlo data is due to the effects of a finite system 
size and
finite temperature. 
Similarly, Monte-Carlo simulations at $T\simeq 0.5
T_c$ and at system sizes around 10, should be able to show 
the
negative slope in $P(q=0,L)$. In the not too far future, it 
should
become possible to perform these simulations.

The fact that $P(q=0,L)$ does not show asymptotic behavior 
even at
$T=T_c/6$ for the system sizes that we have studied, is 
surprising,
and is different from our findings in $d=3$ \cite{us1}. That
$P(q=0,L)$ converges slower towards the asymptotic behavior 
than the
flow of the coupling constant (see Fig.~\ref{flow}), can be 
understood
in the following way: A Parisi overlap value close to zero 
can be
generated by a domain wall excitation. For large system sizes 
and low
temperatures, such an excitation occurs with significant 
weight only
in those samples where a domain wall excitation costs little 
energy.
These are exactly the samples with a small renormalized 
coupling
constant at system size $L$. As the width of the probability
distribution function of the couplings increases with 
$L^\theta$, the
probability for obtaining a small renormalized coupling 
decreases as
$L^{-\theta}$. This is the argument that predicts that 
$P(q=0,L) \sim
L^{-\theta}$. However, for smaller system sizes and higher
temperatures, there are corrections to this argument. Thus, 
even
samples with a renormalized coupling that is not small can 
contribute
to $P(q=0,L)$ by means of large or multiple droplet 
excitations, or of
thermally activated domain walls. For this reason, $P(q=0,L)$ 
can be
expected to converge towards asymptopia slower that the 
coupling
constant itself. Furthermore, as we shall see in the next 
section, the
superposition of domain wall excitations and droplet 
excitations leads
to deviations from simple scaling, which may further slow 
down the
convergence towards asymptotic scaling behavior.

\section{The Link Overlap}

The link overlap gives additional information about the spin 
glass phase that is
not readily seen in the Parisi overlap.  The main qualitative
differences between the Parisi overlap and the link overlap 
are (i)
that flipping all spins in one of the two replicas changes 
the sign of
$q$ but leaves $q_l$ invariant, and (ii) that flipping a 
droplet of
finite size in one of the two replicas changes $q$ by an 
amount
proportional to the volume of the droplet, and $q_l$ by an 
amount
proportional to the surface of the droplet. Thus, the link 
overlap
contains information about the surface area of excitations. 

\begin{figure}
\centerline{\epsfysize=0.7\columnwidth{\epsfbox{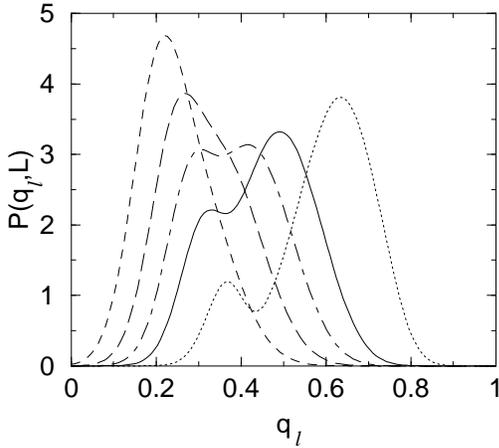}}}
\narrowtext{\caption{$P(q_l,L)$ for $T=0.8T_c$, $0.67T_c$,
      $0.56T_c$, $0.48T_c$, $0.33T_c$ (from left to right), 
with the
      system size $L=4$. }\label{pql2} 
}
\end{figure}

\begin{figure}
  
\centerline{\epsfysize=0.7\columnwidth{\epsfbox{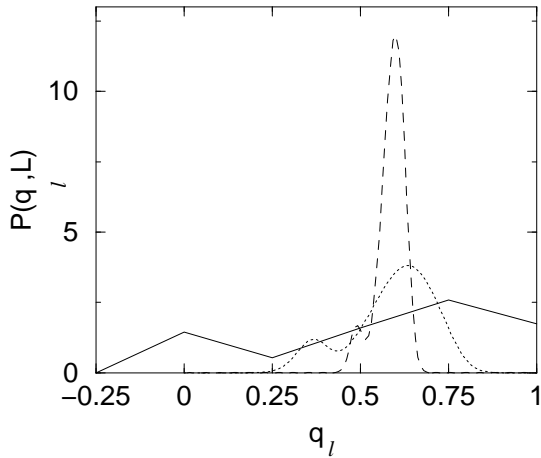}}}
  \narrowtext{\caption{$P(q_l,L)$ for $L=2$, 4, 8 (from 
widest to
      narrowest curve) and with $T=0.33T_c$. 
}\label{pql_1.414} 
}
\end{figure}

First, let us study $P(q_l,L)$ as function of temperature, 
for a given
system size $L=4$. Fig.~\ref{pql2} shows our curves for 
$T=0.8T_c$,
$0.67T_c$, $0.56T_c$, $0.48T_c$, and $0.33T_c$. They appear 
to result
from the superposition of two different peaks, with their 
distance
increasing with decreasing temperature, and the weight 
shifting from
the left peak to the right peak. 
Fig.~\ref{pql_1.414} shows $P(q_l,L)$ for fixed $T=0.33T_c$ 
and for
different $L$. One can see that with increasing system size 
the peaks
move closer together, and the weight of the left-hand peak 
decreases.

These results are similar to what we found in MKA in three 
dimensions
\cite{us4}, however, in four dimensions the peaks are more 
pronounced.
Monte-Carlo simulations of the four-dimensional Ising spin 
glass also show two
peaks for certain system sizes and temperatures \cite{cir93}. 
This feature is
attributed by the authors to RSB. However, as it is also 
present in MKA, there
must be a different explanation. The width of the curves 
shrinks with
increasing system size in \cite{cir93}, just as it does in 
MKA and as is
expected from the droplet picture. If the RSB scenario were 
correct, the width
would go to a finite value in the limit $L\to \infty$.

\begin{figure}
\centerline{\epsfysize=0.7\columnwidth{\epsfbox{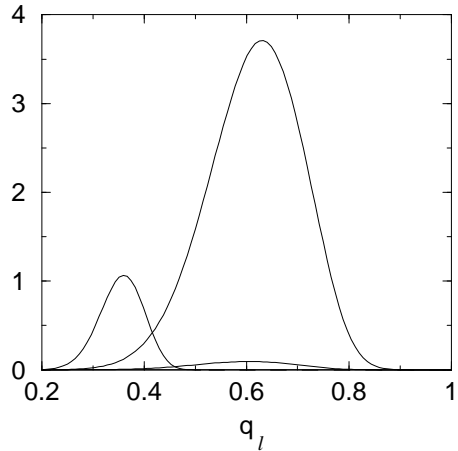}}}
\narrowtext{\caption{Contribution of domain wall excitations 
(left curve) and
droplet excitations (right curve) to $P(q_l,L)$, for $L=4$ 
and $T=0.33T_c$. The
third, flat curve is due to configurations where both 
replicas have a domain
wall. }\label{pql_bc}  }
\end{figure}

In the following we present evidence that the left peak 
corresponds to
configurations where one of the two replicas has a domain 
wall
excitation, and the right peak to configurations where one of 
the two
replicas has a droplet excitation.  In MKA, domain wall 
excitations
involve flipping of one side of the system, including one of 
the two
boundary spins of the hierarchical lattice, while droplet 
excitations
involve flipping of a group of spins in the interior. If the 
sign of
the renormalized coupling is positive (negative), the two 
boundary
spins are parallel (antiparallel) in the ground state. By 
plotting
separately the contributions from configurations with and 
without
flipped boundary spins, we can separate domain wall 
excitations from
droplet excitations. Fig.~\ref{pql_bc} shows the three 
contributions
from configurations where none, one, or both replicas have a 
domain
wall. Clearly, the left peak is due to domain wall 
excitations, and the right
peak to droplet excitations.
Similar curves are obtained for other values of the 
parameters.  We
thus have shown that the qualitative differences between 
droplet and
domain wall excitations are sufficient to explain the 
structure of the
link overlap distribution, and no other low-lying excitations 
like
those invoked by RSB are needed. 

The weight with which domain-wall excitations occur is in agreement
with predictions from the droplet model.
The probability of having a domain wall in a system of size $L$ is according to the droplet picture of the order of 
$$
(T/J)L^{-\theta},$$
which is $\simeq 0.25$ at $T=0.33 T_c$ and
$L=4$, and $\simeq 0.15$ at $T=0.33T_c$ and $L=8$. From our
simulations, we find that the relative weights of domain walls for
these two situations are $\simeq 0.12$ and $\simeq 0.076$, which fits
the droplet picture very well if we include a factor 1/2 in the above
expression. Domain walls become negligible only when the product
$(T/J)L^{-\theta}$ becomes small. In higher dimensions, the critical
value of $T/J$ becomes larger, and for a given relative distance from
the critical point, the weight of domain walls therefore also becomes
larger.  This explains why the effect of domain walls is more visible
in 4 dimensions than in 3 dimensions. However, with increasing system
size, domain walls should become negligible more rapidly in higher
dimensions, due to the larger value of the exponent $\theta$.

\section{Conclusions}
\label{conclusions}

Our results for the Parisi overlap distribution in four 
dimensions
show that there are rather large finite size effects in four
dimensions which give rise to phenomena normally attributed 
to RSB.
The system sizes needed to see the beginning of droplet like 
behavior
within the MKA are larger, and the temperatures are lower, 
than those
studied by Monte Carlo simulations.  However, at temperatures 
not too
far below those studied in Monte Carlo simulations 
($T=0.5T_c$), the
weight of the Parisi overlap distribution function $P(q=0,L)$ 
within
the MKA appears to decrease, albeit with an effective 
exponent
different from the asymptotic value.  Thus, simulations at 
these
temperatures for the Ising spin glass on a cubic lattice 
might resolve
the controversy regarding the nature of the ordered state in 
short
range spin glasses.  However, the MKA is a low dimensional
approximation and it is possible that the system sizes needed 
to see
asymptotic behavior for a hypercubic lattice in four 
dimensions are
different from what is indicated by the MKA. So, any 
comparison of the
MKA with the Monte Carlo data should be taken with a pinch of 
salt.

Recently, a modified droplet picture was suggested by 
Houdayer and
Martin \cite{hou99}, and by Bouchaud \cite{bou99}. Within this 
picture,
excitations on length scales much smaller than the system 
size are
droplet-like, however, there exist large-scale excitations 
that extend
over the entire system and that have a small energy that does 
not
diverge with increasing system size. As we have demonstrated 
within
MKA, the double-peaked curves for the link overlap 
distribution, can
be fully explained in terms of two types of excitations that
contribute to the low-temperature behavior, namely 
domain-wall
excitations and droplet excitations. We therefore believe 
that there
is no need to invoke system-wide low-energy excitations that 
are
more relevant than domain walls.

Finally, the whole field of dynamical studies of spin glasses 
is thought by
many \cite{silvio.pc} to provide a strong reason for 
believing the RSB picture. 
A very recent study of spin glass dynamics on the 
hierarchical lattice 
\cite{fredrico}, on which the MKA is exact, indicates that no 
ageing occurs at 
low temperatures in the response function  whereas in 
Monte-Carlo simulations
on the Edwards-Anderson model and in spin glass experiments  
ageing is seen in
the response function. We suggest that the ageing behavior 
found in Monte-Carlo
simulations and experiment are in fact often dominated by 
critical point
effects,
and not by droplet effects. Indeed we would expect that if 
the simulations of
Ref. \cite{fredrico} were performed at temperatures closer to 
the critical
temperature then ageing effects would be seen in the response 
function, since
near the critical point of even a ferromagnet such ageing 
effects occur
\cite{Godreche}. The reason why experiments and simulations 
on the
Edwards-Anderson model see ageing in the response function   
is that they are
probing time scales that may be less than  the critical time 
scale, which is
given by $$\tau = \tau_0 (\xi/a)^z,$$ with $a$ the lattice 
constant and
$\tau_0$ the characteristic spin-flip time. The dynamical 
critical exponent  $z
\simeq 6$ in 3 dimensions \cite{alan}. Only for droplet 
reversals which take
place on time scales larger than $\tau$ (i.e for reversals of 
droplets whose
linear dimensions exceed $\xi$) will droplet results for the 
dynamics be
appropriate. However, because of the large values of $\xi$ 
down to temperatures
of at least $0.5T_c$ and the large value of $z$, $\tau$ may 
be very large in
the Monte-Carlo simulations and experiments. Thus if $\xi/a$ 
is 100, then
$\tau/\tau_0$ is $10^{12}$, which would make droplet like 
dynamics beyond the
reach of a Monte-Carlo simulation. In practice, most data 
will be in a crossover
regime leading to an apparently temperature dependent  
exponent $z(T)$ (see for
example Ref \cite{komori}).

\acknowledgements We thank A.~P.~Young for discussions and 
for
encouraging us to write this paper. Part of this work was 
performed
when HB and BD were at the Department of Physics, University 
of
Manchester, supported by EPSRC Grants GR/K79307 and 
GR/L38578. BD also
acknowledges support from the Minerva foundation.

\end{multicols} 
\end{document}